\documentclass[manuscript,nonacm]{acmart}

\usepackage{subcaption}
\usepackage{algpseudocode}
\usepackage{algorithm}
\usepackage[frozencache,cachedir=.]{minted}

\algnewcommand\algorithmicforeach{\textbf{for each}}
\algdef{S}[FOR]{ForEach}[1]{\algorithmicforeach\ #1\ \algorithmicdo}


\newcommand{\name}{\textsc{dae4hls}}

\begin{document}

\title{\name{}: Exposing Memory-Level Parallelism for  High-Level Synthesis using Explicit Decoupling}

\author{David Metz}
\email{David.C.Metz@ntnu.no}
\orcid{0000-0001-7103-7968}
\affiliation{%
 \institution{Norwegian University of Science and Technology (NTNU)}
 \streetaddress{Department of Computer Science}
 \city{Trondheim}
 \country{Norway}
 \postcode{7491}
}

\author{Magnus Sj\"alander}
\email{Magnus.Sjalander@ntnu.no}
\orcid{0000-0003-4232-6976}
\affiliation{%
 \institution{Norwegian University of Science and Technology (NTNU)}
 \streetaddress{Department of Computer Science}
 \city{Trondheim}
 \country{Norway}
 \postcode{7491}
}

\begin{abstract}
High-level synthesis (HLS) performs well for simple memory access patterns, such as for sequential accesses that can be turned into bursts, or for memory accesses into small datasets that can be stored in scratchpads.
This limits HLS to accelerating only the low-hanging fruit, where memory-level parallelism is either trivially abundant, due to simple access patterns, or latency is low, due to the small dataset.
Applications with more complex access patterns on large datasets would also benefit from acceleration, and would especially benefit from the reduction in design and verification effort that HLS promises.

In this paper, we present \name{}, a decoupled access-execute (DAE) paradigm for HLS.
We propose a new programming model for explicitly decoupling requests and responses, which unlocks memory-level parallelism that otherwise cannot be automatically provided by a compiler.
We apply the \name{} paradigm to the commercial AMD Vitis HLS toolchain and show that the existing AXI stream and AXI burst interfaces can be repurposed for explicit decoupling.
We further apply the paradigm to a dynamic-HLS framework, which is better suited for handling irregular workloads as compared to statically scheduled HLS.
We show that support for explicit decoupling improves the performance and achieves a total speedup of 10-79$\times$.

\end{abstract}

\maketitle

\section{Introduction}

Irregular memory accesses are common for many data structures, ranging from linked lists to hash tables, and are prevalent in a wide range of applications, such as graph processing and ray-tracing.
Unfortunately, the memory wall~\cite{Sally-memory-wall:CANEWS1995} has made their access latency increasingly costly and difficult to optimize automatically.
CPUs rely on costly speculation to expose memory-level parallelism (MLP) to hide the latency of irregular memory accesses, while being at an inherent disadvantage, in terms of efficiency, compared to dedicated accelerators due to their programmability~\cite{domain-specific-accelerators:CACM2020}.
Dedicated accelerators can be very efficient, but the complexity of designing and verifying them manually makes them prohibitively expensive for many workloads.
High-level synthesis (HLS) promises to solve this, but memory optimizations in HLS tools focus on locality.
Locality-based optimizations work well for regular access patterns, but irregular/random access workloads defeat caches and require large windows of requests to be reordered~\cite{asiatici-moms:PhD2021}.

The key to accelerating applications with irregular memory accesses is to identify available memory-level parallelism and generate enough overlapping memory accesses to hide the access latency.
A new class of memory subsystem optimizes for misses, coalescing, and reordering, rather than focusing on caching~\cite{asiatici-moms:PhD2021}, enabling accelerators to maintain numerous independent requests in flight.
Compilers are increasingly powerful but generally operate at an abstraction level close to machine code, which is generated from a programming language with limited expressiveness of the programmer's intent.
They are required to make conservative assumptions to ensure program correctness.
Compilers, and thus HLS tools, are, therefore, still not sufficiently powerful to extract enough MLP from irregular applications without help.
At the same time, programming frameworks such as the Open Computing Language (OpenCL)~\cite{opencl:V3_0_19} and CUDA~\cite{cuda:12_8} are designed for embarrassingly data-parallel memory access, and do not provide the expressiveness required for controlling individual streams of data dependencies required to extract MLP from irregular memory accesses efficiently.

Manual exposure of hardware mechanisms has historically enabled performance breakthroughs that compilers could not achieve automatically.
This is especially the case when communication and memory accesses are concerned, since they are less predictable than arithmetic instructions.
Even in highly optimized application domains, compilers alone are insufficient.
When the AI space had its \emph{DeepSeek} moment, where a relatively unknown company managed to attain a state-of-the-art model with surprisingly few resources, this was in part enabled by directly using PTX intrinsics rather than fully relying on a higher-level compiler~\cite{deepseek:ARXIV2025}.
While there have been decades of research into auto-vectorization, many performance-critical use cases for SIMD still rely on hand-optimized assembly.
Both of these examples demonstrate that exposing hardware features to expert programmers can unlock new acceleration avenues that can, potentially, be automated later.

\name{} provides a decoupled access-execute paradigm and framework for HLS that enables performance-optimization experts to express the orchestration of memory operations without having to create a detailed implementation and the necessary transformations to take the proposed programming interface into functional circuits (\autoref{sec:dae4hls}).
In \name{}, the programmer identifies memory accesses that benefit from decoupling by leveraging domain knowledge.
\name{} can be used to precisely fetch demand data, e.g., suitable for removing false dependencies (\autoref{sec:false-dependencies}), or to initiate a large number of parallel memory accesses, e.g., suitable for pointer-chasing scenarios in data structures (\autoref{sec:pointer-chasing}).
The \name{} framework leverages explicit decoupling to automatically instantiate request and response logic, enabling multiple requests to be issued and for the data to be consumed once a response is received from the memory subsystem.
Thus, \name{} provides the means to utilize the potential of memory subsystems that support a large number of outstanding requests, without requiring the functionality to be implemented in RTL.

We examine four applications with irregular memory accesses to showcase the power of explicit decoupling using both repurposed burst constructs in the commercial AMD Vitis HLS~\cite{vitis:V2024_2} toolchain (\autoref{sec:realization-vitis}) and an optimized dynamic HLS framework (\autoref{sec:realization-rhls}).
Our results show that \name{} produces accelerators with significantly improved throughput (10$\times$ to 79$\times$) while reducing resource consumption (\autoref{sec:results}).

\section{Background}
\label{sec:background}

In this section, we present prefetching, streaming, and decoupled access execute with their main benefits and drawbacks at a conceptual level, recent advancements in HLS and accelerator memory subsystems for massive memory-level parallelism (MLP), and dynamic HLS's benefits over conventional HLS.

\begin{figure}[t]
    \centering
    \begin{subfigure}[b]{0.3\columnwidth}
        \centering
        \includegraphics[width=\textwidth]{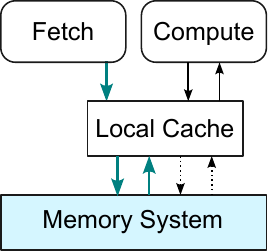}
        \caption{Prefetch}
        \label{fig:overview-prefetch}
    \end{subfigure}
    \hfill
    \begin{subfigure}[b]{0.3\columnwidth}
        \centering
        \includegraphics[width=\textwidth]{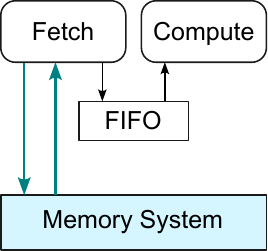}
        \caption{Stream}
        \label{fig:overview-stream}
    \end{subfigure}
    \hfill
    \begin{subfigure}[b]{0.3\columnwidth}
        \centering
        \includegraphics[width=\textwidth]{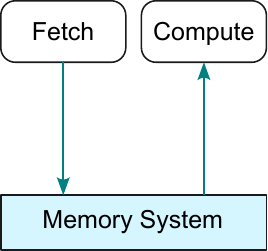}
        \caption{Decouple}
        \label{fig:overview-decouple}
    \end{subfigure}
    \caption{Illustration showing the general principle behind (a)~prefetching, (b)~streaming, and (c)~decoupling.}
    \label{fig:overview}
\end{figure}

\subsection{Decoupled Access Execute}
\label{sec:background-dae}

While decoupled access execute (DAE) \cite{dae:CANEWS1982} is related to pre\-fetching, in that they both aim to improve MLP and timeliness, there are distinct differences.
Prefetch requests are, in essence, requests without a consumer for the response, as illustrated in \autoref{fig:overview-prefetch}.
Instead, they are used to provide data to a future request at a closer point in the memory hierarchy, and thus become more timely.
This means that prefetching can be speculative and does not need to be perfect to function correctly, even though that would be desirable for performance reasons.
However, prefetching requires address calculations to be performed twice: first for the (pre)\emph{Fetch}, and later for the request from \emph{Compute} to the \emph{Local Cache}, as shown in the figure.
Since prefetching is speculative, it does not guarantee that data is kept in the cache or has been prefetched in the first place.
This is indicated by the dotted set of connections from the \emph{Local Cache} to the \emph{Memory System}.
Prefetching, by its nature, increases the total number of requests to the memory subsystem, requiring more cache accesses to be handled.

Another related concept to DAE is streaming, in which data between computation stages is placed in a FIFO for efficient consumption at a later stage.
An example where memory requests are performed in the first stage
is illustrated in \autoref{fig:overview-stream}.
Streaming requires a precise number of requests, since the number of writes and reads to the FIFO must be equal to avoid deadlocks.

Decoupled access execute (DAE) reduces memory access latency by performing memory requests early, and depends on the memory system to independently provide the requested data to a separate compute kernel, as illustrated in \autoref{fig:overview-decouple}.
Decoupling is a specialized form of streaming that eliminates the FIFO and utilizes the memory subsystem instead.
With decoupling, similarly to streaming, the number of requests made is not increased compared to the baseline application.
Furthermore, the requests are less data-dependent than in streaming, which improves timeliness.
However, decoupling depends on a more specialized memory system, and not all applications are suitable for streaming or decoupling, i.e., decoupling is not suitable for all memory-access patterns.

\name{} provides explicit streaming and DAE primitives for fine-grained control over memory accesses in HLS.

\subsection{The Miss Optimized Memory Subsystem}
\label{sec:moms}

While caches help to save time and energy when there is temporal or spatial locality to exploit, even an optimally organized cache will have a lot of misses when dealing with random accesses to a large amount of memory.
In this scenario, the best a memory subsystem can do is to combine accesses to the same regions, eliminating bank and row changes.
Instead of focusing on storing data, the subsystem can focus on storing the metadata of a large window of accesses, enabling large-scale reordering and coalescing.

Recent developments in memory systems for complex accelerators lend themselves well to the decoupled access-execute (DAE) concept~\cite{dae:CANEWS1982}.
The miss-optimized memory system (MOMS)~\cite{asiatici-moms:PhD2021} is a new class of memory subsystem targeting FPGAs, which focuses on large-scale request coalescing and reordering rather than caching.
Instead of primarily storing cached data, the MOMS focuses on storing request metadata, enabling significantly more in-flight requests than a conventional cache with miss status handling registers (MSHRs) can handle.
Once a cache line arrives from DRAM, it is stored only temporarily to generate responses for already outstanding requests to that cache line.
The large number of in-flight requests enables reordering to optimize the access pattern to DRAM banks and rows.
This can, for example, be combined with modern FPGA platforms that provide ample DRAM bandwidth at a low per-bit energy cost using high-bandwidth memory (HBM), and enable efficient handling of random accesses across the entire DRAM.
The MOMS complements DAE by enabling large request windows and reordering, which decoupled accelerators can exploit to achieve high throughput on large irregular workloads.

\subsection{Memory Optimizations in HLS}

While HLS has proven effective for regular workloads, its locality-based optimizations fail for irregular access patterns that require thread-like memory concurrency.
State-of-the-art HLS tools have support for specifying multiple bursts and collecting their responses in order, but do not facilitate thread-like behavior with multiple outstanding requests.
Thus, conventional HLS struggles to provide the required memory-level parallelism (MLP) needed for efficient acceleration of irregular applications.
Several works in the HLS and the accelerator domain achieve speedups or save power by eliminating memory accesses, caching them, or localizing them to scratchpads~\cite{scalehls:HPCA2022, stream-hls:FPGA2025}.
This makes sense, since moving data, and especially accessing DRAM, is expensive.
However, applications with irregular accesses to large datasets do not lend themselves to these kinds of optimizations.
Memory optimizations for dynamic HLS have focused on dynamic memory disambiguation between loads and stores~\cite{ooo-lq:TECS2017, straight-to-the-queue:FPGA2023, lsq-sizing:ICFPT2022, r-hls:ICCAD2024}.
While these techniques help unlock performance for applications with aliasing loads and stores, helping some irregular applications, they do not unlock MLP for dependent or control-dependent loads.
Unlike these techniques \name{} exposes MLP directly, enabling efficient and performant HLS acceleration of irregular accesses.

\subsection{Dynamic HLS and R-HLS}

Dynamically scheduled HLS is a growing branch of high-level synthesis~\cite{dynamically-scheduled-hls:FPGA2018, dynamatic:FPGA2020, fluid:ASYNC2021, fast-tokens:FPL2022, inter-block-scheduling:FPL2022, parallel-control-flow:TRETS2023, straight-to-the-queue:FPGA2023, r-hls:ICCAD2024, crush:ASPLOS2025}.
Unlike static HLS, where scheduling is performed at compile time and the accelerator is controlled by a state machine, dynamic HLS accelerators are dataflow circuits that schedule operations dynamically using handshaking.
While variable memory latency has been highlighted as an area where dynamic HLS could provide benefits, previous dynamic HLS work has focused on accelerators with low-latency memory accesses to SRAM-based scratchpads, making them unsuitable comparison points.

R-HLS~\cite{r-hls:ICCAD2024} is a dynamic HLS framework based on the regionalized value state dependence graph (RVSDG)~\cite{rvsdg:TECS2020} intermediate representation.
RVSDG is a dataflow-based intermediate representation that models control flow as regions and captures memory dependencies using state edges.
R-HLS is an RVSDG dialect that models information required for HLS.
For example, R-HLS exploits state edges for distributed memory disambiguation and synchronization~\cite{r-hls:ICCAD2024}.

\section{\name{} --- Explicit Decoupling}
\label{sec:dae4hls}

The goal of this paper is to evaluate what kind of performance explicit decoupling can unlock for suitable applications.
We focus on providing interfaces, which we employ for manual optimization, rather than trying to define compiler passes that optimize these particular applications automatically.
These interfaces and exploration provide a foundation that can be used for further optimizations.
As described in \autoref{sec:optimizing-applications}, this enables us to exploit application characteristics that are at best implicitly encoded and at worst data-dependent and not deducible from the program alone.

We utilize two concepts to implement explicit decoupling---streams and decoupled loads.
Streams can already be found in static HLS tools and enable values to be transported between different locations in a program without having to follow program semantics~\cite{behavioral-soc:DAC2021, stream-hls:FPGA2025, vitis:V2024_2}.
We use streams as a baseline to compare against, to transport state between iterations for pointer chasing, and as an inspiration of our decoupled load primitive.

\begin{listing}[t]
    \begin{minted}[fontsize=\footnotesize, escapeinside=@@]{c}
//Access
for (uint32_t i = 0; i < table_elements; i++)
{
  load_result = table[i];
  @\fcolorbox{blue}{white}{stream\_enq(stream\_channel, load\_result)}@;
  @\fcolorbox{red}{white}{decouple\_request(load\_channel, &table[i])}@;
}
//Execute
for (uint32_t i = 0; i < table_elements; i++)
{
  sum1 += @\fcolorbox{blue}{white}{stream\_deq(stream\_channel, capacity)}@;
  sum2 += @\fcolorbox{red}{white}{decouple\_response(load\_channel, capacity))}@;
}
    \end{minted}
    \caption{Decouple and stream semantics.}
    \label{code:decouple-semantics}
\end{listing}

\subsection{In-Order Streams}
\label{sec:in-order-streams}

We encode point-to-point streams as function calls.
As shown in \autoref{code:decouple-semantics}, marked blue, these consist of a \emph{stream\_enq} and a \emph{stream\_deq}.
The connection between the two is symbolized by them sharing a channel constant, in this case \emph{stream\_channel}.
The call to \emph{stream\_enq} also takes a value, in this case \emph{load\_result}.
The enqueued values are returned by \emph{stream\_deq} in the same order they were enqueued.

The implicit dependency between the \emph{stream\_enq} and \emph{stream\_deq} calls enables the two loops to be instantiated as separate execution units that run in parallel.
For a more complex example with a longer running execute loop, the \emph{Access} loop can run ahead and fetch the \emph{table[]} data, while the \emph{Execute} loop consumes the data as soon as an iteration is completed.

The capacity of the FIFO that connects the enqueue and the dequeue point is given by the second argument to \emph{stream\_deq}, i.e., \emph{capacity}.
The number of enqueue calls during execution must match the number of dequeue calls, ensuring that there are no values remaining in the stream after execution finishes.
Since the two loops can run in parallel, the stream does not need to be bigger than the number of table elements in this example, which it would need to be if the \emph{Access} loop had to finish before the \emph{Execute} loop starts.
Instead, the FIFO size controls how many iterations ahead of the \emph{Execute} loop the \emph{Access} loop can run.

\subsection{Why Decoupled Loads and not Just Streams?}
\label{sec:why-decoupled}

A decoupled load is similar to a stream, but instead of a value, it takes a pointer, i.e., an address to be loaded, as shown in red in \autoref{code:decouple-semantics}.
Conceptually, the load producing \emph{load\_result} and the stream in \autoref{code:decouple-semantics} together encode the same behavior as the decoupled load.
This is the scenario visualized in \autoref{fig:overview-stream} and \autoref{fig:overview-decouple}.
Combining them into one and relying on the memory subsystem to supply the response removes the required FIFO for the stream and reduces the pipeline depth of \emph{Access}.
For example, when the load has a latency of 100 cycles, \emph{Access} requires a pipeline depth of at least 100, to be able to wait for the load response and enqueue it into the stream while maintaining throughput.
This applies to both static and dynamic HLS.
The decoupled load is also beneficial from a memory ordering perspective, since it cleanly splits when the request can be made, and when the response arrives.
This is explored in more detail in \autoref{sec:realization-rhls}.
A decouple operation could be inferred from a stream consuming a load by a compiler, and we explored this approach in Vitis.
We choose to distinguish them for clarity.

\begin{listing}[t]
\begin{minipage}{0.49\textwidth}
    \begin{minted}[fontsize=\footnotesize, escapeinside=@@]{c}
void spmv(float val[NNZ], int32_t cols[NNZ],
          int32_t rows[N+1], float vec[N], float out[N])
{






  // Original kernel
  for (int i = 0; i < N; i++)
  {
    float sum = 0;
    for (int j = @\fcolorbox{yellow}{white}{rows[i]}@; j < @\fcolorbox{yellow}{white}{rows[i+1]}@; j++)
    {
      float Si = @\fcolorbox{green}{white}{val[j]}@ *
                 @\fcolorbox{red}{white}{vec[\colorbox{cyan}{cols[j]}]}@;
      sum = sum + Si;
    }
    out[i] = sum;
  }
}
    \end{minted}
\end{minipage}
\hfill
\begin{minipage}{0.49\textwidth}
    \begin{minted}[fontsize=\footnotesize, escapeinside=@@]{c}
void spmv(float val[NNZ], int32_t cols[NNZ],
          int32_t rows[N+1], float vec[N], float out[N])
{
  // Access
  for (int j = 0; j < NNZ; j++)
  {
    @\fcolorbox{green}{white}{decouple\_request(val\_channel, &val[j])}@;
    @\fcolorbox{red}{white}{decouple\_request(vec\_channel, &vec[\colorbox{cyan}{cols[j]}]))}@;
  }
  // Execute
  for (int i = 0; i < N; i++)
  {
    float sum = 0;
    for (int j = @\fcolorbox{yellow}{white}{rows[i]}@; j < @\fcolorbox{yellow}{white}{rows[i+1]}@; j++)
    {
      float Si = @\fcolorbox{green}{white}{decouple\_response(val\_channel)}@ *
                 @\fcolorbox{red}{white}{decouple\_response(vec\_channel)}@;
      sum = sum + Si;
    }
    out[i] = sum;
  }
}
    \end{minted}
\end{minipage}
\caption{SPMV with (left)~compressed sparse row input matrix, based on the implementation in Machsuite~\cite{machsuite:IISWC2014}, and (right)~decoupled version that can run ahead.}
\label{fig:spmv}
\end{listing}

\section{Optimizing Irregular Applications}
\label{sec:optimizing-applications}

In this section, we show how the proposed explicit decoupling interface can be used to improve the
performance of irregular applications by removing false data dependencies and enabling parallel pointer chasing.

\subsection{Removing False Data Dependencies}
\label{sec:false-dependencies}

For two of the examined workloads, SPMV and Mergesort, decoupling can be used to remove false data dependencies of loads.
We designate data dependencies that occur in the program that could be avoided using more expressive semantics as false data dependencies.

\textbf{SPMV} implements sparse matrix-vector multiplication of a dense vector and a sparse matrix in compressed sparse-row (CSR) format, producing a dense output.
The CSR format consists of three arrays containing values, column indices, and row delimiters.
Each non-zero in the matrix is stored as a combination of a value and a column index, sorted by row.

An implementation of CSR SPMV is shown in \autoref{fig:spmv} (left).
Each row has a start and end index into the values and columns, with the end index of a row being the start index of the following row.
In this implementation, the accesses to \emph{val}, \emph{vec}, and \emph{cols} are dependent on an access to \emph{rows}.
$rows[0] = 0$ and $rows[N] = NNZ$ (number of non-zeros), with row values in between increasing monotonically.
This means that each index in \emph{val} and \emph{cols} is accessed once sequentially.

\begin{listing}[t]
\begin{minipage}{0.49\textwidth}
    \begin{minted}[fontsize=\footnotesize, escapeinside=@@]{c}
void merge(const uint32_t *table, uint32_t *result,
           uint32_t i_left, uint32_t i_right, uint32_t i_end)
{






  // Original kernel
  uint32_t @\colorbox{red}{table\_i}, \colorbox{green}{table\_j}@;
  uint32_t i = i_left;
  uint32_t j = i_right;
  for (uint32_t k = i_left; k < i_end; ++k)
  {
    // Read each element

    @\colorbox{red}{table\_i}@ = table[i];


    @\colorbox{green}{table\_j}@ = table[j];

    // Select the smallest element
    if ((@\colorbox{red}{table\_i}@ <= @\colorbox{green}{table\_j}@ || j >= i_end) && i < i_right)
    {

      result[k] = @\colorbox{red}{table\_i}@;
      i++;
    }
    else
    {

      result[k] = @\colorbox{green}{table\_j}@;
      j++;
    }
  }
}
    \end{minted}
\end{minipage}
\hfill
\begin{minipage}{0.49\textwidth}
    \begin{minted}[fontsize=\footnotesize, escapeinside=@@]{c}
void merge(const uint32_t *table, uint32_t *result,
           uint32_t i_left, uint32_t i_right, uint32_t i_end)
{
  // Access
  for (uint32_t i_req = i_left; i_req < i_right; i_req++)
    @\fcolorbox{red}{white}{decouple\_request(i\_channel, &table[i\_req]);}@
  for (uint32_t j_req = i_right; j_req < i_end; j_req++)
    @\fcolorbox{green}{white}{decouple\_request(j\_channel, &table[j\_req]);}@
  // Execute
  bool @\colorbox{yellow}{update\_table\_i}@ = true, update_table_j = true;
  uint32_t @\colorbox{red}{table\_i}, \colorbox{green}{table\_j}@;
  uint32_t i = i_left;
  uint32_t j = i_right;
  for (uint32_t k = i_left; k < i_end; ++k)
  {
    // Read each element only once
    if (@\colorbox{yellow}{update\_table\_i}@ && i < i_right)
      @\colorbox{red}{table\_i}@ = @\fcolorbox{red}{white}{decouple\_response(i\_channel)}@;
    @\colorbox{yellow}{update\_table\_i}@ = false;
    if (update_table_j && j < i_end)
      @\colorbox{green}{table\_j}@ = @\fcolorbox{green}{white}{decouple\_response(j\_channel)}@;
    update_table_j = false;
    // Select the smallest element
    if ((@\colorbox{red}{table\_i}@ <= @\colorbox{green}{table\_j}@ ||   j >= i_end) && i < i_right)
    {
      @\colorbox{yellow}{update\_table\_i}@ = true;
      result[k] = @\colorbox{red}{table\_i}@;
      i++;
    }
    else
    {
      update_table_j = true;
      result[k] = @\colorbox{green}{table\_j}@;
      j++;
    }
  }
}
    \end{minted}
\end{minipage}
   \caption{Merge sort with (left)~the original merge function and (right)~the decoupled version.}
    \label{fig:mergesort}
\end{listing}

\autoref{fig:spmv} (right) demonstrates an altered version of SPMV that leverages decoupling to exploit this property.
Now \emph{val}, \emph{vec}, and \emph{cols} are accessed sequentially, without depending on accesses to \emph{rows}.

\textbf{Mergesort} implements a bottom-up version of the merge-sort sorting algorithm.
Initially, it partitions an array of length $N$ that is being sorted into $N$ separate runs of length~1.
Since each run has only a single element, each run is sorted.
Pairs of these runs are then merged, resulting in sorted runs of twice the length.
This merging process is repeated until there is only a single sorted run.
The \emph{merge} function of Mergesort is shown in \autoref{fig:mergesort} (left).
The first run uses the index \emph{i} and the second run the index \emph{j}.
In this version, loads are dependent on a load-dependent comparison, since the increment of \emph{i} and \emph{j} is dependent on a comparison of \emph{table[i]} and \emph{table[j]}.

\autoref{fig:mergesort} (right) demonstrates an altered version of Mergesort using decoupling.
As we know that both runs are accessed sequentially, we can issue decouple requests for all elements in them and consume the responses sequentially.
To load each value only once, \emph{table\_i} (red) and \emph{table\_j} (green), which hold the first element of a run that has not yet been consumed, are introduced.
\emph{update\_table\_i} (yellow) is used to ensure each element is loaded only once, removing redundant memory accesses.
At the beginning of each iteration, \emph{update\_table\_i} is used to decide if \emph{table\_i} should be updated using a decouple response.
This is the case in the first iteration, and later on, if \emph{update\_table\_i} gets set when \emph{i} gets incremented.
The decouple request loops run in parallel to the merge loop, so that the decoupled loads only require enough buffer capacity to hide the memory latency, but not enough to hold the entire run.

Implementing the requests as part of the merge, as demonstrated here, does not always unlock the full performance potential, since runs are very short at the beginning of merge sort.
What can be done instead is to run these loops ahead for the following runs as well, outside of the merge function.

After all pairs of runs have been merged, they are stored in \emph{result}.
Since HLS tools commonly do not support swapping pointers, the values have to be copied back to \emph{table} before the next iteration.
Alternatively, if the final sorted output being stored in either \emph{result} or \emph{table} is acceptable, a second version of the sort iteration can be instantiated in place of the copy loop.
This increases resource consumption but decreases the number of loads and stores, and thus the runtime, from $2 n \log_2 n - n$ to $n \log_2 n$, i.e., almost a factor of two.

\subsection{Enabling Parallel Pointer Chasing}
\label{sec:pointer-chasing}

Two of the workloads, Hashtable and Binsearch, are essentially pointer-chasing workloads.
Hashtable looks up integer values in a hash map with index-based overflow buckets.
Binsearch looks up integer values in a sorted array using binary search.
These and similar workloads are common in databases.
In pointer-chasing workloads, the result of a load determines the address of the next load.
This makes pure pointer-chasing inherently serial, with no MLP, and thus at first glance a poor fit for DAE.
However, when looking up multiple values, MLP can be extracted by following multiple pointer chains in an overlapping manner.
To maximize MLP, as many values should be looked up in parallel as the memory latency in cycles.

\begin{listing}[t]
\begin{minipage}{0.49\textwidth}
    \begin{minted}[fontsize=\footnotesize, escapeinside=@@]{c}
for (uint32_t i = 0; i < tableElements; ++i)
{
  @\colorbox{yellow}{state}@ = get_initial_state(i);
  while (true)
  {
    next_addr = get_addr(@\colorbox{yellow}{state}@);
    load_result = @\fcolorbox{red}{white}{*next\_addr}@;
    @\colorbox{yellow}{state}@ = update_state(@\colorbox{yellow}{state}@, load_result);
    if (end_reached(@\colorbox{yellow}{state}@)
    {
      write_result(@\colorbox{yellow}{state}@);
      break;
    }
  }
}
    \end{minted}
\end{minipage}
\hfill
\begin{minipage}{0.49\textwidth}
    \begin{minted}[fontsize=\footnotesize, escapeinside=@@]{c}
uint32_t @\colorbox{green}{rif = 0}@; // requests in flight
uint32_t table_i = 0;
for (uint32_t finished = 0; finished < table_elements;)
{
  if (@\colorbox{green}{rif < RIF}@ && table_i < table_elements)
  {
    @\colorbox{yellow}{state}@ = get_initial_state(table_i++);
    @\colorbox{green}{rif++}@;
  }
  else
  {
    load_result = @\fcolorbox{red}{white}{decouple\_response(load\_channel, RIF+1))}@;
    @\colorbox{yellow}{state}@ = @\fcolorbox{blue}{white}{stream\_deq(state\_stream, RIF+1)}@;
    @\colorbox{yellow}{state}@ = update_state(@\colorbox{yellow}{state}@, load_result);
    if (end_reached(@\colorbox{yellow}{state}@)
    {
      write_result(@\colorbox{yellow}{state}@);
      finished++;
      @\colorbox{green}{rif--}@;
      continue;
    }
  }
  next_addr = get_addr(@\colorbox{yellow}{state}@);
  @\fcolorbox{red}{white}{decouple\_request(load\_channel, next\_addr)}@;
  @\fcolorbox{blue}{white}{stream\_enq(state\_stream, \colorbox{yellow}{state})}@;
}
    \end{minted}
\end{minipage}
    \caption{Programming model for (left)~simple sequential pointer chasing and (right)~parallel pointer chasing.}
    \label{fig:pointer-chasing}
\end{listing}

\autoref{fig:pointer-chasing} (left) shows a simple generalization of pointer-chasing, following the pointer chains one-by-one sequentially.
\autoref{fig:pointer-chasing} (right) outlines a decoupled parallel approach for these workloads that overlaps pointer chains.
At a higher abstraction level, the approach can be thought of as request-in-flight (RIF) threads, processing one pointer chain each, and sharing execution resources in a round-robin fashion.
Each thread has a \emph{state} (yellow) that is passed through a stream (blue) alongside the decoupled long-latency request (red).

The for loop runs until all pointer chains have been followed to their end and are \emph{finished}.
The number of pointer chains currently being followed is \emph{rif} (green), with the maximum amount being controlled by \emph{RIF}.
If the maximum is not yet reached, and there are elements left to process, a new pointer chain is started.
This is achieved by initializing the state based on the table entry that is being looked up in \emph{get\_initial\_state}.
Otherwise, the oldest request's response is processed.
When the end of a pointer chain is reached (\emph{end\_reached}), \emph{rif} is decremented, so a new pointer chain can be followed.
The loop is continued, so no request is issued this iteration.

\begin{listing}[t]
    \begin{minted}[fontsize=\footnotesize, escapeinside=@@]{c}
for (uint32_t i = 0; i < table_elements; i = chunk_end)
{
  chunk_end = i + CHUNK_SIZE;
  for (uint32_t j = 0; j < chase_iterations; j++)
  {
    for (uint32_t k = i; k < chunk_end; ++k)
    {
      if (first_iteration)
        @\colorbox{yellow}{state}@ = get_initial_state(k);
      else
      {
        load_result = @\fcolorbox{red}{white}{decouple\_response(load\_channel, CHUNK\_SIZE+1))}@;
        @\colorbox{yellow}{state}@ = @\fcolorbox{blue}{white}{stream\_deq(state\_stream,  CHUNK\_SIZE+1)}@;
        @\colorbox{yellow}{state}@ = update_state(@\colorbox{yellow}{state}@, load_result);
      }
      if (last_iteration)
        write_result(@\colorbox{yellow}{state}@);
      else
      {
        next_addr = get_addr(@\colorbox{yellow}{state}@);
        @\fcolorbox{red}{white}{decouple\_request(load\_channel, next\_addr)}@;
        @\fcolorbox{blue}{white}{stream\_enq(state\_stream, \colorbox{yellow}{state})}@;
      }
    }
  }
}
    \end{minted}
    \caption{Fixed length parallel pointer chasing.}
    \label{fig:bounded_pointer_chasing}
\end{listing}

An alternative approach, when the pointer chain length is bounded, is shown in \autoref{fig:bounded_pointer_chasing}.
Similar to the first approach in \autoref{fig:pointer-chasing} (right), the resolution of multiple, in this case, ideal fixed-length pointer chains is parallelized.
However, unlike the first approach, the overlap is less fine-grained and instead performed in chunks that operate in lock-step.
First, the initial state of all pointer chains in the chunk is generated.
Then, the first pointer is followed for all the chains, and all pointers are followed until the last pointer of all chains in the chunk.
This approach generates and writes results in order and could thus simplify the memory subsystem for outputs.
This works for binsearch, but has a penalty in terms of cycle count and amount of loads performed, because the search for an element cannot finish early if an element is found.
The number of loads could be reduced to the level of the first approach by making the \emph{decouple\_request} and \emph{decouple\_response} calls conditional based on \emph{state}, at the cost of circuit complexity.
This would, however, not affect the higher number of iterations required.

\section{Realizing Explicit Decoupling}
\label{sec:realization}

In this section, we present how the explicit decoupling described in \autoref{sec:dae4hls} can be integrated into a commercial HLS toolchain and a dynamic HLS framework.

\subsection{Correctness}
\label{sec:correctness}

For decoupling circuits to function correctly, all requests require a corresponding response.
This must be ensured by the programmer, and is one of the differences between decoupling and prefetching.
For SPMV, the original code and the decoupled version in \autoref{fig:spmv} are only equivalent as long as the CSR matrix data structure is correct, i.e., $rows[i] <= rows[i+1]$ and $rows[N+1]-rows[0] == NNZ$.
We argue that this is an acceptable trade-off, since a malformed CSR matrix would, regardless, not result in a correct SPMV computation, and, in general, malformed data can easily cause deadlocks or infinite loops in both software and accelerators.
The decoupled implementation of Mergesort in \autoref{fig:mergesort} ensures that both \emph{i\_channel} and \emph{j\_channel} have the correct number of response accesses based on a careful construction of the conditions that update \emph{update\_table\_i} and \emph{update\_table\_j}.
$j >= i\_end$ ensures that a value from \emph{i\_channel} is used when no values are left for \emph{j\_channel} and $i < i\_right$ ensures the equivalent with the channels swapped.
For the parallel pointer chasing in \autoref{fig:pointer-chasing}, the \emph{continue} statement ensures that no \texttt{decouple\_request()} and \texttt{stream\_enq()} calls are made at the end of a pointer chain.
For the fixed-length version in \autoref{fig:bounded_pointer_chasing}, the same is ensured by the \emph{if (last\_iteration)} statement.

\subsection{AMD Vitis}
\label{sec:realization-vitis}

We showcase how explicit decoupling can be integrated into the AMD Vitis toolchain~\cite{vitis:V2024_2}.
Vitis can be instructed to run loops in parallel by placing them in separate sub-functions and applying the \emph{dataflow} pragma.
The \texttt{stream\_enq()} and \texttt{stream\_deq()} can be implemented as \emph{hls::stream}s in Vitis.
The \texttt{decouple\_request()} and \texttt{decouple\_response()} can be implemented using the \emph{hls::burst\_maxi} interfaces, with bursts of length one.
The input pointer is replaced with a \emph{burst\_maxi} object.
Multiple \emph{burst\_maxi} objects can be connected to the same AXI interface, but their requests may not overlap.
This means that each pointer can only be accessed by one request/response pair at a time, making this implementation unsuitable for the proposed optimization of Mergesort in \autoref{sec:false-dependencies}.

Creating static schedules for variable latency memory accesses requires either \textit{i)} assuming the worst-case latency, causing unnecessary overhead and requiring a sufficient number of outstanding memory requests, or \textit{ii)} using an average latency, resulting in lost performance when the access latency is longer.
As will be shown in \autoref{sec:results}, this significantly limits the performance improvements and increases the hardware resource usage for the evaluated benchmarks.

\subsection{Explicit Decoupling with R-HLS}
\label{sec:realization-rhls}

Dynamic HLS inherently adapts to variable access latencies, making it suitable for irregular memory accesses that result in variable latencies.
Here, we showcase how explicit decoupling can be integrated in a state-of-the-art dynamic HLS framework called R-HLS~\cite{r-hls:ICCAD2024}.

To preserve program correctness when introducing explicit decoupling, we need to ensure that potentially aliasing loads and stores, i.e., loads and stores that may access the same memory location, are ordered, and all operations have finished.
R-HLS uses state edges for distributed memory disambiguation and synchronization~\cite{r-hls:ICCAD2024}.
If, for example, a function being turned into an accelerator has no return value and instead writes its results to memory, a state edge is what ensures that the accelerator does not terminate early and waits for all memory operations to complete.
This means that, to synchronize with a later, potentially aliasing, store, the state edge ensures that the store waits for the load(s) to complete.

\begin{figure}[t]
    \centering
    \begin{minipage}{0.75\textwidth}
    \begin{subfigure}[b]{0.55\textwidth}
        \centering
        \includegraphics[width=\textwidth]{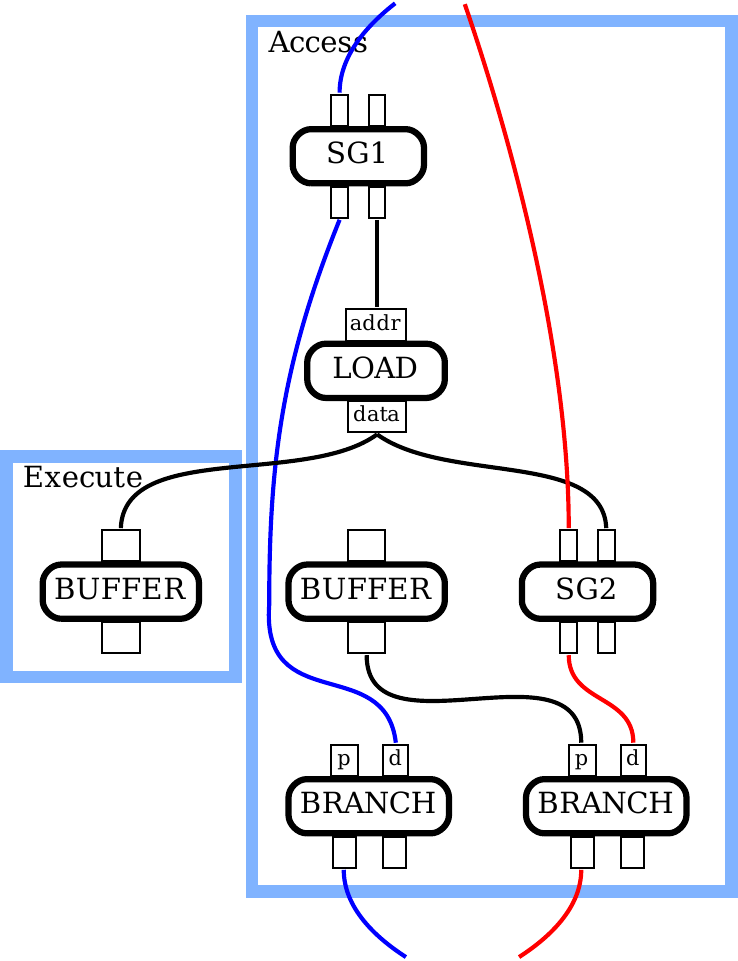}
        \caption{Stream}
        \label{fig:state-stream}
    \end{subfigure}
    \hfill
    \begin{subfigure}[b]{0.3\textwidth}
        \centering
        \includegraphics[width=\textwidth]{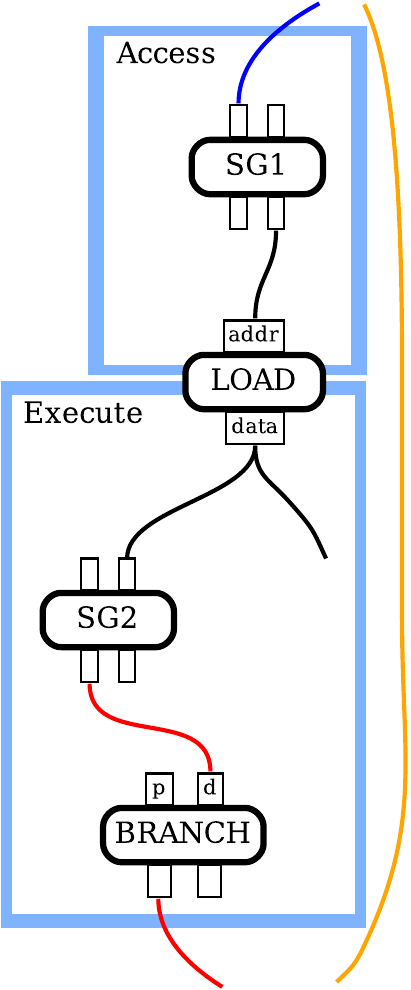}
        \caption{Decouple}
        \label{fig:state-decouple}
    \end{subfigure}
    \caption{State edges for stream and decouple in \autoref{code:decouple-semantics}.}
    \label{fig:state-stream-decouple}
    \end{minipage}
\end{figure}

\textbf{Conventional streaming: } \autoref{fig:state-stream} shows what is required when using a separate load and stream, as in \autoref{code:decouple-semantics}.
State gates synchronize state edges with data and only let either through when both are present.
A state edge with a state gate is used to synchronize when the load may start (blue, \emph{SG1}), by gating its address, and when it is finished (red, \emph{SG2}), by waiting for its data response.
R-HLS uses branches at the bottom of loops to control when values loop back up or exit the loop.
As discussed in \autoref{sec:why-decoupled}, if the memory latency is 100, \emph{Access} needs to have a pipeline depth of at least 100.
This means the buffer controlling the branch needs to hold at least 100 elements.
Additionally, a FIFO buffer for the stream is required in \emph{Execute}, since otherwise \emph{Execute} might hinder the progression of \emph{Access}, because the \emph{SG2} state gate and branch will only receive new loaded values if \emph{Execute} also consumes them.

\textbf{Explicit decoupling:} \autoref{fig:state-decouple} shows the equivalent when using a decoupled load.
Now the decoupled target loop, \emph{Execute}, is responsible for the state edge signaling that the operation has finished.
Since there is no benefit for parts of \emph{Execute} to run ahead of the decoupled response, \emph{Execute}'s pipeline depth is one, so no buffer is needed for the branch, and no buffer is required for the loaded value.

The load straddles the boarder between \emph{Access} and \emph{Execute}.
The load still receives the response with decoupling, to ensure it does not issue more loads than it has capacity for, which could result in a deadlock.
An example where such a deadlock could occur without this mechanism is Mergesort, where two different decoupled accesses share a memory port, and one of them could exhaust all request capacity, leaving none for the other access.

Furthermore, we use different synchronization mechanisms for decoupled loads than for normal loads.
While normal loads use the distributed disambiguation circuit described by Metz et al.~\cite{r-hls:ICCAD2024}, decoupled loads forgo address queues, since they provide little benefit in most scenarios where decoupling is applicable.
Instead of two state-edges, we utilize three.
The blue state edge still controls when the load can execute, but ends at the state gate \emph{SG1}.
The red state edge now starts at the state gate \emph{SG2}.
The red and blue state edges are implicitly connected through the load, and together form a continuous state edge.

However, if \emph{table\_elements} were zero, the load would not be executed, and no synchronization could occur. 
This scenario is handled by the orange state edge.
The three state edges are handled separately in the largest region possible.
If there are no stores that may alias, they are split apart at the beginning of the accelerator execution and merged in the end.
If there are stores that may alias, they are split for the largest region possible and merged before the store is encountered.

\subsection{Long-Latency Memory}
\label{sec:rhls-ll}

Previous work on dynamic HLS focused on low-latency BRAM memory.
Moving to a more complex memory subsystem with longer latency, connected via AXI, requires several changes, which we outline for R-HLS.
To ensure that the generated accelerator does not deadlock, loads can only perform as many requests as they have buffer capacity.
Otherwise, a scenario where one load issues too many requests, such that it cannot accept all responses from a shared memory response port, blocking responses to other loads and stalling progress permanently, can occur.
For BRAM with single-cycle latency, this can be avoided by allowing load requests to be issued if a single-entry buffer is empty.
The general topic of resource sharing in dynamic HLS has been studied by Xu and Josipovic~\cite{crush:ASPLOS2025}.

In BRAM, stores are observable to loads in the same cycle, or the cycle after they have been performed, depending on BRAM configuration.
This enables stores to be counted as observable a single cycle after they are sent off.
In contrast to BRAMs, e.g., when using an AXI memory subsystem, stores can only be counted as observable once a corresponding response to the store request has been received via the write response channel.
To ensure there can be no deadlocks of the memory response, buffer capacity must be provided for this response as well.
Previously, R-HLS used a single state-edge to synchronize stores~\cite{r-hls:ICCAD2024}.
This meant an initiation interval of one was only achievable if there were no other cycle delays for the state edge in a loop, since the store released the state edge one cycle after it issued its request.
However, with increased memory latencies, this is no longer feasible, since the response may arrive many cycles later, increasing the initiation interval.

The solution to this problem is to view the synchronization of the store not as one synchronization point, or state edge, but as two.
One to control when the operation can start, and one to control when it has finished, or is observable.
The distributed disambiguation scheme proposed by Metz et al.~\cite{r-hls:ICCAD2024} already applies this concept to loads, with a separate state edge per load controlling when a load can start, and a shared state edge synchronizing load results with stores.

Since the load-store interactions of the workloads studied in this paper are relatively simple, with no load-store aliasing for Hashtable, Binsearch, and SPMV, and no aliasing within inner loops, but only across loops for Mergesort and Multi-SPMV, there is a more resource-efficient alternative to separate state-edges.
As long as they are performed in-order, a store node can perform multiple stores to the same address without having to worry about ordering.
This is because memory operations are assigned static AXI IDs, and store operations are guaranteed to be performed in-order for an AXI ID, with reordering only possible across IDs.

\begin{figure}[t]
    \centering
    \begin{minipage}{0.75\textwidth}
    \begin{subfigure}[b]{0.35\textwidth}
        \centering
        \includegraphics[width=\textwidth]{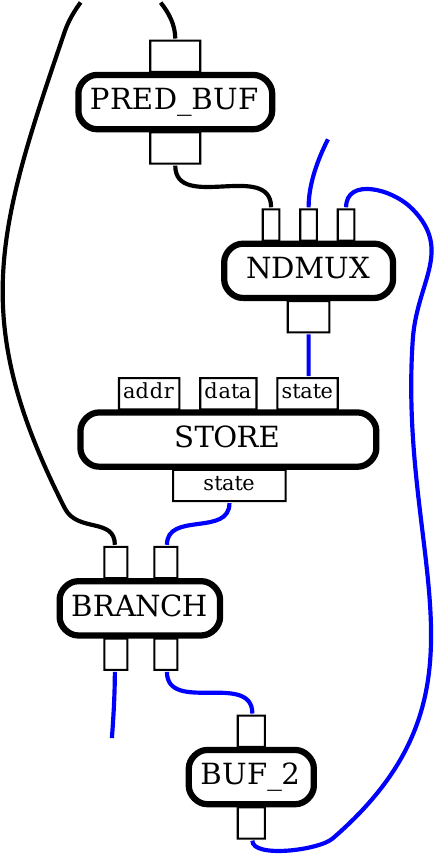}
        \caption{Conventional back-edge}
        \label{fig:store-normal}
    \end{subfigure}
    \hfill
    \begin{subfigure}[b]{0.35\textwidth}
        \centering
        \includegraphics[width=\textwidth]{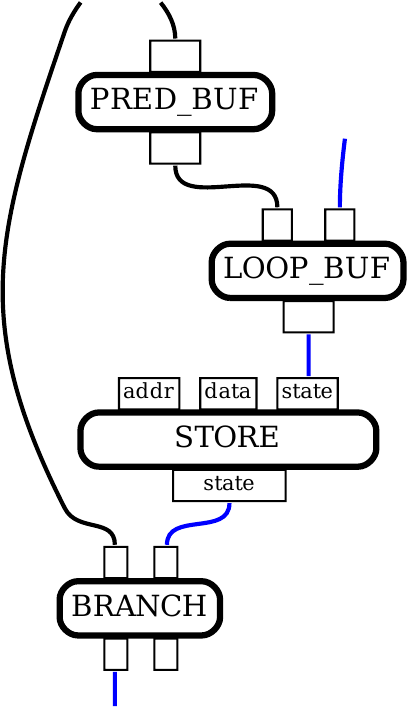}
        \caption{Using a loop buffer}
        \label{fig:store-lcb}
    \end{subfigure}
    \caption{Store state edge.}
    \end{minipage}
\end{figure}

This means that when there is only a single operation on a state-edge in a loop, the operations can start without having to wait for the back-edge.
A conventional state-edge for a store in a loop is shown in \autoref{fig:store-normal}.
\autoref{fig:store-lcb} shows an optimized version without the back edge, which uses a loop buffer instead of a non-discarding multiplexor at the top of the loop.
This enables stores to be issued as soon as the loop input and loop predicate are available, enabling the execution, i.e., period between request and response, of subsequent store requests to overlap.
The branch operation at the bottom still guarantees that the last store has received its response before the state edge is passed on from the loop.
This scheme can also be applied to nested loops, as long as there is only one memory operation on the state edge.

Previously, R-HLS used shift-register-based buffers.
These do not scale to the buffer sizes needed for long memory latencies, so memory-based buffers are employed instead.

\section{Methodology}
\label{sec:methodology}

For static HLS, we generate accelerators using AMD Vitis 2024.1~\cite{vitis:V2024_2}.
\emph{Vitis} uses Vitis-generated accelerators without decoupling.
\emph{Vitis Decoupled} uses explicit decoupling as described in \autoref{sec:realization-vitis}.
Vitis does not allow multiple parallel processes to access the same AXI interface, severely limiting the achievable speedup in Mergesort.

For dynamic HLS, we augment R-HLS~\cite{r-hls:ICCAD2024} with the changes necessary to effectively support long-latency memory operations and decoupling described in \autoref{sec:rhls-ll} and \autoref{sec:realization-rhls}.
Furthermore, we employ a new approach for the placement of transparent buffers.
This approach calculates the cycle depth of all operations and places buffers on the inputs of operations if they have a lower cycle depth than the other inputs.
Since this can lead to overestimates in buffer size, we employ a profile-guided optimization of buffer sizes.
Previously, R-HLS accelerators used BRAM memories.
Instead, we equip accelerators with one AXI interface per pointer argument.
To decouple the critical path of the accelerator from the memory subsystem, we place buffers on the AXI interfaces.

We evaluate three different configurations for R-HLS.
\emph{R-HLS} and \emph{R-HLS Stream} use the memory disambiguation by Metz et al.~\cite{r-hls:ICCAD2024}, with the changes to accommodate long latency memory operations described in \autoref{sec:rhls-ll}.
\emph{R-HLS} does not use decoupling, while \emph{R-HLS Stream} uses a combination of loads and streams to approximate decoupling.
\emph{R-HLS Stream} does not have results for \texttt{mergesort}, since the memory disambiguation scheme creates a dependency between the two fetch loops, leading to a deadlock.
\emph{R-HLS Decoupled} implements full decoupling.

We use Verilator to simulate memory accesses with a latency of 100 cycles for both reads and writes in a unified manner for the R-HLS and Vitis accelerator.
To gather additional results with a more realistic memory subsystem, we integrate accelerators running at $100\,MHz$ with the Miss Optimized Memory Subsystem (MOMS)~\cite{asiatici-moms:PhD2021} configured with $128\,KiB$ of cache, $3\times512$ entry hash tables, and a $512\,bit$ external AXI interface with $64$ outstanding reads.
The external AXI interface is connected to a $333\,MHz$ DDR3 memory interface simulated using DRAMSim2~\cite{dramsim2:CAL2011}.
Since the MOMS is a read-only cache without coherency, we only provide results for the subset of benchmarks compatible with this restriction.
The other benchmarks would require flushing of the MOMS between read and write phases, or a more advanced memory subsystem.
Critical path and resource utilization are evaluated by synthesizing the accelerators using Vivado 2024.1 with a clock period of $10\,ns$.
Both the Vitis and R-HLS generated accelerators instantiate Vivado floating-point IPs for SPMV.

We evaluate across seven different benchmarks.
\texttt{binsearch} and \texttt{binsearch\_for} implement binary search for multiple elements in a sorted array, and differ in that the former has an early exit condition, with the latter using a constant number of iterations.
\texttt{hashtable} implements a hash table lookup for multiple elements, with the hash table using separate chaining, i.e., maintaining a linked list of values with equal hashes, for each hash bucket.
\texttt{mergesort} implements bottom-up merge sort of integers.
\texttt{mergesort\_opt} instantiates a second search iteration instead of the copy loop.
\texttt{spmv} implements vector-matrix-multiplication of a dense vector and a sparse matrix in CSR format.
\texttt{multispmv} implements multiple iterations of \texttt{spmv}, with the output of \texttt{spmv} being scaled and copied back to the input vector after each iteration.

We scaled input sizes for applications using the MOMS so that the irregularly accessed data would exceed the $128\,KB$ cache capacity of the MOMS.
This leaves some opportunities for reuse, without making things trivial to cache.
\texttt{binsearch} and \texttt{binsearch\_for} look up $1\,000$ elements in a sorted array of $1\,234\,567\times4B$ elements.
\texttt{hashtable} follows $1\,024$ $\approx16$ long pointer chains through an array of $65\,536\times16B$ entries.
\texttt{spmv} performs matrix-vector multiplication of a $1\,024\times16\,777\,216$ element matrix with $17\,221$ non-zero element and a $16\,777\,216\times4B$ dense vector.
\emph{mergesort} sorts $234\times4B$ numbers.
\texttt{multispmv} performs ten matrix-vector multiplications of a $128\times128$ element matrix with $1\,639$ non-zero element and a $128\times4B$ dense vector.

\section{Results}
\label{sec:results}

\begin {table*}[t]
\caption{Performance with Path given in nanoseconds and Time in microseconds.}
\label{tab:performance}
\resizebox{\linewidth}{!}{
\begin{tabular}{l|rrr|rrrr|rrrr|rrrr|rrrr}
\toprule
 & \multicolumn{3}{c|}{Vitis} & \multicolumn{4}{c|}{Vitis Decoupled} & \multicolumn{4}{c|}{R-HLS} & \multicolumn{4}{c|}{R-HLS Stream} & \multicolumn{4}{c}{R-HLS Decoupled} \\
 & Cycles & Path & Time
 & Cycles & Path & Time & Speedup
 & Cycles & Path & Time & Speedup
 & Cycles & Path & Time & Speedup
 & Cycles & Path & Time & Speedup \\
\midrule
binsearch & 2\ 298\ 439 & 7.62 & 17\ 507.21 & 65\ 091 & 6.88 & 447.7 & 39.11 & 2\ 039\ 174 & 9.44 & 19\ 243.69 & 0.91 & 21\ 364 & 12.51 & 267.22 & 65.52 & 21\ 354 & 12.16 & 259.71 & 67.41 \\
binsearch\_for & 2\ 357\ 243 & 7.05 & 16\ 620.92 & 83\ 937 & 8.77 & 736.21 & 22.58 & 2\ 163\ 106 & 9.29 & 20\ 101.74 & 0.83 & 22\ 230 & 9.14 & 203.25 & 81.78 & 22\ 206 & 9.53 & 211.67 & 78.52 \\
hashtable & 1\ 953\ 903 & 6.57 & 12\ 829.33 & 53\ 887 & 5.87 & 316.26 & 40.57 & 1\ 687\ 760 & 9.08 & 15\ 331.61 & 0.84 & 19\ 292 & 10.8 & 208.35 & 61.57 & 19\ 086 & 10.47 & 199.85 & 64.19 \\
mergesort & 259\ 157 & 7.95 & 2\ 060.04 & 145\ 423 & 8.99 & 1\ 307.64 & 1.58 & 199\ 862 & 9.83 & 1\ 964.24 & 1.05 &  &  &  &  & 7\ 038 & 9.44 & 66.44 & 31.01 \\
mergesort\_opt &  &  &  &  &  &  &  &  &  &  &  &  &  &  &  & 3\ 960 & 9.6 & 38 &  \\
multispmv & 348\ 343 & 8.53 & 2\ 969.62 & 60\ 243 & 8.54 & 514.48 & 5.77 & 71\ 214 & 13.84 & 985.25 & 3.01 & 32\ 218 & 13.68 & 440.84 & 6.74 & 21\ 904 & 13.46 & 294.78 & 10.07 \\
spmv & 286\ 379 & 8.62 & 2\ 467.73 & 55\ 071 & 8.58 & 472.51 & 5.22 & 18\ 644 & 14 & 261.03 & 9.45 & 17\ 532 & 13.42 & 235.31 & 10.49 & 17\ 530 & 13.56 & 237.69 & 10.38 \\
\bottomrule
\end{tabular}

}
\end{table*}

\begin {table*}[t]
\caption{Resource usage.}
\label{tab:resources}
\resizebox{\linewidth}{!}{
\begin{tabular}{l|cccc|cccc|cccc|cccc|cccc}
\toprule
 & \multicolumn{4}{c|}{Vitis} & \multicolumn{4}{c|}{Vitis Decoupled} & \multicolumn{4}{c|}{R-HLS} & \multicolumn{4}{c}{R-HLS Stream} & \multicolumn{4}{c}{R-HLS Decoupled} \\
 & LUTs & FFs & BRAMs & DSPs & LUTs & FFs & BRAMs & DSPs & LUTs & FFs & BRAMs & DSPs & LUTs & FFs & BRAMs & DSPs & LUTs & FFs & BRAMs & DSPs \\
\midrule
binsearch & 3\ 390 & 3\ 987 & 1 & 0 & 3\ 600 & 3\ 828 & 3 & 0 & 2\ 956 & 1\ 647 & 2 & 0 & 4\ 511 & 1\ 691 & 1 & 0 & 4\ 226 & 1\ 605 & 1 & 0 \\
binsearch\_for & 3\ 491 & 4\ 053 & 1 & 0 & 4\ 055 & 4\ 412 & 2 & 0 & 3\ 036 & 1\ 900 & 4 & 0 & 3\ 446 & 2\ 087 & 2 & 0 & 3\ 174 & 1\ 979 & 2 & 0 \\
hashtable & 3\ 192 & 4\ 325 & 2 & 0 & 3\ 284 & 3\ 956 & 3 & 0 & 1\ 864 & 1\ 536 & 2 & 0 & 3\ 710 & 1\ 597 & 5 & 0 & 3\ 476 & 1\ 494 & 3 & 0 \\
mergesort & 4\ 825 & 5\ 308 & 1 & 0 & 5\ 401 & 6\ 052 & 4 & 0 & 7\ 460 & 4\ 653 & 2 & 0 &  &  &  &  & 4\ 978 & 2\ 228 & 1 & 0 \\
mergesort\_opt &  &  &  &  &  &  &  &  &  &  &  &  &  &  &  &  & 7\ 743 & 3\ 074 & 2 & 0 \\
multispmv & 8\ 405 & 9\ 172 & 4 & 5 & 8\ 763 & 9\ 758 & 4 & 8 & 8\ 443 & 5\ 687 & 3 & 6 & 8\ 284 & 5\ 292 & 2 & 6 & 6\ 224 & 3\ 362 & 2 & 6 \\
spmv & 5\ 914 & 6\ 479 & 2 & 5 & 6\ 205 & 6\ 973 & 2 & 5 & 4\ 455 & 2\ 903 & 5 & 4 & 3\ 724 & 2\ 189 & 2 & 4 & 3\ 393 & 2\ 004 & 2 & 4 \\
\bottomrule
\end{tabular}
}
\end{table*}

\begin {table*}[t]
\caption{Performance with MOMS.}
\label{tab:performance_moms}
\resizebox{\linewidth}{!}{
\begin{tabular}{l|rrr|rrrr|rrrr|rrrr|rrrr}
\toprule
 & \multicolumn{3}{c|}{Vitis} & \multicolumn{4}{c|}{Vitis Decoupled} & \multicolumn{4}{c|}{R-HLS} & \multicolumn{4}{c|}{R-HLS Stream} & \multicolumn{4}{c}{R-HLS Decoupled} \\
 & cycles & path & time & cycles & path & time & speedup & cycles & path & time & speedup & cycles & path & time & speedup & cycles & path & time & speedup \\
\midrule
binsearch & 2\ 239\ 063 & 7.62 & 17\ 054.94 & 65\ 011 & 6.88 & 447.15 & 38.14 & 677\ 274 & 9.44 & 6\ 391.43 & 2.67 & 23\ 310 & 12.51 & 291.56 & 58.5 & 23\ 302 & 12.16 & 283.4 & 60.18 \\
binsearch\_for & 2\ 294\ 243 & 7.05 & 16\ 176.71 & 83\ 937 & 8.77 & 736.21 & 21.97 & 701\ 472 & 9.29 & 6\ 518.78 & 2.48 & 25\ 938 & 9.14 & 237.15 & 68.21 & 25\ 928 & 9.53 & 247.15 & 65.45 \\
hashtable & 1\ 904\ 751 & 6.57 & 12\ 506.6 & 53\ 887 & 5.87 & 316.26 & 39.54 & 1\ 008\ 246 & 9.08 & 9\ 158.91 & 1.37 & 18\ 782 & 10.8 & 202.85 & 61.66 & 18\ 716 & 10.47 & 195.98 & 63.82 \\
spmv & 283\ 829 & 8.62 & 2\ 445.75 & 55\ 037 & 8.58 & 472.22 & 5.18 & 29\ 918 & 14 & 418.88 & 5.84 & 29\ 864 & 13.42 & 400.83 & 6.1 & 29\ 732 & 13.56 & 403.14 & 6.07 \\
\bottomrule
\end{tabular}
}
\end{table*}

The performance of the different accelerators is presented in \autoref{tab:performance}, and their resource consumption in \autoref{tab:resources}.
Compared to \emph{Vitis}, \emph{R-HLS Decoupled} achieves a speedup of 10.07$\times$ -- 78.52$\times$,
while compared to \emph{Vitis Decoupled} the speedup is reduced to 1.58$\times$ -- 19.68$\times$.
Resource consumption is consistently lower for \emph{R-HLS Decoupled}, except for \texttt{binsearch}, where it requires more lookup tables (LUTs).
Flip-flops (FFs) are generally lower for R-HLS by a factor of at least two.
R-HLS designs have a longer critical path, which is offset by lower cycle counts.
Vitis's static scheduling is unable to schedule an initiation interval of one for critical loops of these workloads, resulting in consistently higher cycle counts.
The clearest advantage for \emph{R-HLS Decoupled} is for \texttt{mergesort}, where competing methods run into limitations with simultaneous memory accesses and streams, and have to fall back to less performant alternatives.
\texttt{mergesort\_opt} provides a 74.8\% speedup over \texttt{mergesort}, at a 55.5\% LUT, 40\% FF, and 100\% BRAM increase.
This seems like a worthwhile trade-off, considering memory traffic is also reduced, and as discussed below, the performance difference exceeds 90\% for larger array sizes.

For \emph{R-HLS Decoupled}, \texttt{binsearch\_for} seems preferable over \texttt{binsearch}, since it achieves a considerably lower critical path that offsets the slightly higher cycle count.
Resource-wise \texttt{binsearch\_for} requires more BRAMs and FFs, i.e., buffer storage, while requiring fewer LUTs.
Compared to \emph{R-HLS Stream}, \emph{R-HLS Decoupled} has similar speedup for \texttt{binsearch}, \texttt{binsearch\_for}, \texttt{hashtable}, and \texttt{spmv}, with \emph{R-HLS Stream} introducing a bottleneck in \texttt{multispmv} and deadlocking in \texttt{mergesort}.
\emph{R-HLS Decoupled} has consistently lower resource consumption, making it the preferred solution.

\begin{figure}[t]
    \centering
    \includegraphics[width=0.5\columnwidth]{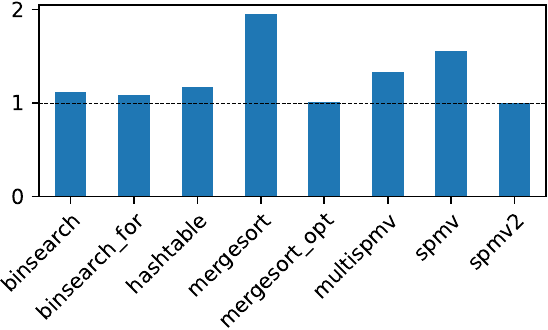}
    \caption{Overhead in cycle count for \name{} over "golden" reference.}
    \label{fig:golden}
\end{figure}

To further evaluate the cycle counts achieved by \name{}, we compare against a "golden" reference, representing a theoretical minimal number of cycles, given the constraint that only one request can be made per pointer argument per cycle.
The golden reference takes only loads to the array being accessed irregularly or data-dependent loads into account, and gives results that would be achieved with no memory latency.
It thus provides a measure of how well long access latency is hidden by \name{}.
For some of the benchmarks, the data set sizes were scaled up compared to \autoref{tab:performance}, since larger datasets amortize the initial startup latency over a longer computation time.
The results are shown in \autoref{fig:golden}.
For the pointer chasing benchmarks \texttt{binsearch}, \texttt{binsearch\_for}, and \texttt{hashtable}, \name{} achieves 11.9\%, 8.6\%, and 17.6\% more cycles than the golden model.
The golden model uses the same cycle count for \texttt{mergesort} and \texttt{mergesort\_opt}, which does not take into account the copy loop of \texttt{mergesort}.
This results in 95.4\% more cycles for \texttt{mergesort}, but only 1.3\% more cycles for \texttt{mergesort\_opt}, close to optimal sorting.
For \texttt{multispmv}, the copy and scale loop is ignored by the golden model, leading to 33.7\% more cycles.
This would be alleviated by a denser or bigger matrix.
For SPMV, we provide two different results, \texttt{spmv}, a relatively sparse matrix at 55.3\% overhead, and \texttt{spmv2}, a denser matrix at 0.3\% overhead.
The number of cycles for the golden model is the number of non-zero elements in the matrix.
For sparser matrices, loading empty rows and storing their results penalize \name{}.

\autoref{tab:performance_moms} presents additional results using a MOMS and simulated DRAM instead of an abstracted memory model with a fixed 100-cycle latency.
The MOMS is only used for the irregular accesses. The other accesses are sequential and assume a fixed 100-cycle latency.
\emph{Vitis} sees cycle counts decrease slightly across workloads, as the average memory access latency is lower, but its static scheduling prevents it from taking advantage of this most of the time.
\emph{Vitis Decoupled} is largely unaffected, since it does not generate enough requests to get back-pressured, but is constrained by its schedule and not memory latency.
Except for \texttt{spmv}, where it was already close to full throughput, \emph{R-HLS} benefits significantly from the reduced latency, demonstrating the advantage of dynamic HLS in this scenario.
\emph{R-HLS Decoupled} has a decreased cycle count for \texttt{hashtable}, presumably due to reduced latency in the end, when only a few pointer chains remain, and MLP is reduced.
For the other workloads \emph{R-HLS Decoupled} sees increased cycle counts, due to back-pressure preventing full throughput.
\emph{R-HLS Stream} behaves analogous to \emph{R-HLS}.
These numbers demonstrate that \name{} performs well under real-world conditions involving variable latency and back-pressure.

\section{Related Work}
\label{sec:related-work}

The concept of separating address generation and data consumption has been frequently explored in the literature.
The term Decoupled Access/Execute was coined by Smith~\cite{dae:CANEWS1982}, for a processor that was split into an access and execute portion.
Pellauer et al.~\cite{buffets:ASPLOS2019} introduce a taxonomy of data orchestration, coined the term Explicit Decoupled Data Orchestration (EDDO), and introduce Buffets, a storage idiom that combines a local scratchpad with state-machine-based DMA, which aims to replace conventional memory hierarchies.
Usui et al.~\cite{usui_takamaeda-yamazaki:ARC2023} take the concept of EDDO and apply automatic decoupling to HLS by using EDDO-based data structures.
There have been several framework proposals for HLS tools performing automated access/execute decoupling~\cite{cheng_wawrzynek:FPT2014, chen_suh:MICRO2016, charitopoulos:CF2018}.
While these can unlock some of the performance available through decoupling, there are inherent limitations to them, such as not being able to overcome false data dependencies.

Nowatzki et al.~\cite{stream-dataflow:ISCA2017} propose Stream-Dataflow Acceleration, which, with its combination of streamed accesses and dataflow computation, resembles the execution model of dynamic HLS with decoupling, but changes the programming paradigm and lists merge sort as unsuitable due to fine-grained data-dependent loads.
Johannes de Fine Licht et al.~\cite{licht+:TPDS2021} describe a process called Memory Access Extraction, in which they separate memory accesses and computation.
However, they target this optimization towards better scheduling and larger access bursts for better bandwidth utilization, rather than unlocking MLP.
AMD Vitis~\cite{vitis:V2024_2} provides the AXI Burst interface, primarily targeted at enabling longer sequential access bursts, which allows for separation of load and store addresses and data.
While it can be repurposed for decoupling, it is not intended or optimized for small independent requests, but large continuous bursts.
A limitation it faces is that only one interface is available per pointer argument, making it unsuitable for merge sort.
Shane T. Fleming and David B. Thomas exploit the concept of Runahead Execution~\cite{runahead-hls:FCCM2017} by automatically creating program slices that perform memory accesses.
Their work does not support both reading and writing to the same pointer, restricting the set of applications that can be targeted.
Suhail Basalama and Jason Cong present Automatic Dataflow Acceleration~\cite{stream-hls:FPGA2025} for HLS, but their technique is limited to affine kernels with perfectly nested loops and focuses on cross-kernel communication.

\section{Conclusion}

\name{} introduces explicit decoupling semantics and demonstrates its potential to unlock memory-level parallelism~(MLP), especially in the context of dynamic HLS.
The proposed programming model and framework provides the means for optimization experts to express the inherent MLP of an application and to automatically generate functional hardware.
For an optimized version of merge sort, a benchmark previously deemed unsuitable for decoupled access execute~(DAE)~\cite{stream-dataflow:ISCA2017}, an optimized version of R-HLS employing \name{} achieves a 54.21x speedup over the Vitis baseline at less than double the resources, while requiring only 1.3\% more cycles than the theoretical optimum.
This demonstrates \name{}'s ability to push HLS to more diverse applications,
and can act as the first step towards automatic optimization by providing a target interface and framework for automatic RTL generation.

\bibliographystyle{ACM-Reference-Format}
\bibliography{refs}

\end{document}